\RequirePackage{fix-cm}
\documentclass[epjc3]{svjour3} 
\usepackage[a4paper,left=2.5cm,right=2.5cm,top=2.5cm,bottom=2.5cm]{geometry}

\RequirePackage{color}
\RequirePackage{graphicx}
\usepackage{setspace}
\usepackage{amsmath}
\usepackage{upgreek}
\usepackage{lineno}
\usepackage[normalem]{ulem}
\usepackage{hyperref}  
\usepackage{xcolor}
\usepackage{siunitx}
\usepackage[export]{adjustbox}

\journalname{Eur. Phys. J. C}

\usepackage{soul}

\begin{document}

\title{Measurement of the electric potential and the magnetic field in the shifted analysing plane of the KATRIN experiment}

\titlerunning{Measurement of SAP fields in KATRIN}        

\thankstext{e1}{Corresponding author: lokhov@kit.edu}

\authorrunning{KATRIN collaboration} 

\normalsize{
\author{KATRIN Collaboration: M.~Aker\thanksref{iap}
\and
D.~Batzler\thanksref{iap}
\and
A.~Beglarian\thanksref{ipe}
\and
J.~Behrens\thanksref{iap}
\and
J.~Beisenk\"{o}tter\thanksref{muenster}
\and
M.~Biassoni\thanksref{infnbicocca}
\and
B.~Bieringer\thanksref{muenster}
\and
Y.~Biondi\thanksref{iap}
\and
F.~Block\thanksref{iap}
\and
S.~Bobien\thanksref{itep}
% ...
}
}

\institute{Institute for Astroparticle Physics~(IAP), Karlsruhe Institute of Technology~(KIT), Hermann-von-Helmholtz-Platz 1, 76344 Eggenstein-Leopoldshafen, Germany\label{iap}
\and
Institute for Data Processing and Electronics~(IPE), Karlsruhe Institute of Technology~(KIT), Hermann-von-Helmholtz-Platz 1, 76344 Eggenstein-Leopoldshafen, Germany\label{ipe}
\and
Institute for Nuclear Physics, University of M\"{u}nster, Wilhelm-Klemm-Str.~9, 48149 M\"{u}nster, Germany\label{muenster}
\and
Istituto Nazionale di Fisica Nucleare (INFN) -- Sezione di Milano-Bicocca, Piazza della Scienza 3, 20126 Milano, Italy\label{infnbicocca}
\and
Institute for Technical Physics~(ITEP), Karlsruhe Institute of Technology~(KIT), Hermann-von-Helmholtz-Platz 1, 76344 Eggenstein-Leopoldshafen, Germany\label{itep}
% ...
\and
Institute for Nuclear Research of Russian Academy of Sciences, 60th October Anniversary Prospect 7a, 117312 Moscow, Russia\footnote{Institutional status in the KATRIN Collaboration has been suspended since February 24, 2022}\label{inr}
}
\date{Received: date / Accepted: date}

\maketitle

\begin{abstract}
The projected sensitivity of the effective electron neutrino-mass measurement with the KATRIN experiment is below 0.3\,eV (\SI{90}{\percent}\,CL) after five years of data acquisition. The sensitivity is affected by the increased rate of the background electrons from KATRIN's main spectrometer. A special shifted-analysing-plane (SAP) configuration was developed to reduce this background by a factor of two. The complex layout of electromagnetic fields in the SAP configuration requires a robust method of estimating these fields. We present in this paper a dedicated calibration measurement of the fields using conversion electrons of gaseous $^\mathrm{83m}$Kr, which enables the neutrino-mass measurements in the SAP configuration.

\keywords{neutrino mass \and KATRIN \and background \and krypton-83m \and electromagnetic field}
\end{abstract}

\large
\section{Introduction}
\label{sec:1}
Neutrinos are the only remaining fundamental particles of the standard model, whose absolute mass is known to be non-zero but has not been determined yet. An upper limit on the sum of the neutrino masses can be set to below \SI{100}{\milli\electronvolt} (\SI{95}{\percent} credible interval) by analysis of the cosmological data~\cite{Planck:2018vyg,Tristram:2023haj,DESI:2024mwx}. Another set of constraints of the order of $\mathcal{O}(100)$~meV is provided by searches for the neutrinoless double beta decay~\cite{Dolinski:2019nrj,CUORE:2024ikf,GERDA:2023wbr,KamLAND-Zen:2024eml}, which could be sensitive to the mass of Majorana neutrinos but require a direct lepton-number violation. In contrast to the above two approaches, direct neutrino-mass measurements using the kinematics of weak decays do not suffer from strong model dependencies~\cite{Formaggio:2021nfz,Lokhov:2022zfn}. 

The  Karlsruhe Tritium Neutrino Experiment (KATRIN) aims to measure the effective mass of the electron antineutrino with a sensitivity below \SI{0.3}{\electronvolt} at \SI{90}{\percent} confidence level~(CL). With its first two science runs, KATRIN reached sub-eV sensitivity to the neutrino mass and placed an upper limit of \SI{0.8}{\electronvolt} (\SI{90}{\percent} CL)~\cite{KATRIN:2019yun,KATRIN:2021uub}. Recently, a new upper limit of \SI{0.45}{\electronvolt} (\SI{90}{\percent} CL) based on the first five science runs, with \num{259} measurement days, was reported~\cite{Aker:2024drp}.
The sensitivity of KATRIN is achieved via the high statistics measurement of the tritium $\upbeta$-decay spectrum near its endpoint with eV energy resolution and excellent control of systematic uncertainties~\cite{KATRIN:2021fgc}.
The ultimate statistical sensitivity of KATRIN depends strongly on the background rate. The measured background level of KATRIN was found to be more than an order of magnitude higher than the design value of 10~mcps (millicounts per second). The main sources of the KATRIN background were identified, investigated and mitigated (see Sec.~\ref{sec:3} for more details). However, the remaining component, associated with the highly excited atoms in the volume of the main spectrometer, persists. 

A new configuration of the electromagnetic fields in the KATRIN main spectrometer, shifted analysing plane (SAP), was proposed as a method to reduce the spectrometer background by a factor of two while preserving the energy resolution~\cite{Lokhov:2022iag}.
The SAP configuration has a more complex, less homogeneous distribution of the electromagnetic fields, compared to the nominal symmetric configuration~\cite{KATRIN:2021dfa}. Since the simulation of such a configuration may not be fully reliable due to imperfect knowledge of the setup geometry, an independent \textit{in situ} measurement of the fields is required. Note that the term ``electromagnetic field'' is used throughout the paper for the magnetic field and the electric potential, which are the quantities of interest.

In this paper, we introduce a method of measuring the fields and transmission properties of the main spectrometer using conversion electrons of $^\mathrm{83m}$Kr. The presented results were used in the recent neutrino mass release by KATRIN~\cite{Aker:2024drp}.

The paper has the following structure. In section~\ref{sec:2}, a brief introduction to the KATRIN experimental setup and measurement principle is presented with an emphasis on the transmission properties of the spectrometer. Section~\ref{sec:3} gives a brief overview of the KATRIN background sources and describes the shifted-analysing-plane configuration. In section~\ref{sec:4}, the principle of measuring the electromagnetic fields in the main spectrometer with gaseous $^\mathrm{83m}$Kr is presented.
In sections~\ref{sec:5} and \ref{sec:6} we describe the analysis of the K-32 and N$_{2,3}$-32 conversion electron lines of $^\mathrm{83m}$Kr, the corresponding systematic uncertainties and the results. In section~\ref{sec:7}, we discuss the impact of the SAP configuration on the neutrino-mass measurements and the analysis strategy, as well as the resulting systematic uncertainties related to the SAP measurement configuration.
The summary and the conclusions are given in section~\ref{sec:8}.

\section{KATRIN setup and measurement principle}
\label{sec:2}
In the KATRIN experiment, electrons are produced in tritium $\upbeta$ decay inside the windowless gaseous tritium source (WGTS) and in a uniform magnetic field. The electrons are guided by a magnetic field towards the spectrometer~\cite{DesignRept2004,KATRIN:2021dfa}. In the transport section the flow of tritium is reduced by 14 orders of magnitude via differential and cryogenic pumping~\cite{Marsteller:2020tgj,Rottele:2023cgk}, preventing tritium from reaching the spectrometers, decaying there and increasing the background. The tandem spectrometers filter the electrons by their kinetic energy, transmitting the high-energy ones to the detector. The KATRIN focal-plane detector, a Si PIN diode segmented into 148 pixels~\cite{Amsbaugh:2014uca}, counts the electrons, which pass the electric-retarding-potential threshold, measuring an integrated energy spectrum. The KATRIN spectrometers are electrostatic filters combined with magnetic adiabatic collimation (MAC-E-filter)~\cite{P_Kruit_1983,LOBASHEV1985305,PICARD1992345}. 
Let us consider the working principle of the KATRIN main spectrometer in more detail to see how the electromagnetic fields enter the description of the measured spectrum~\cite{Kleesiek:2018mel}.

\begin{figure*}[h]
  \frame{\includegraphics[width=0.85\textwidth,center]{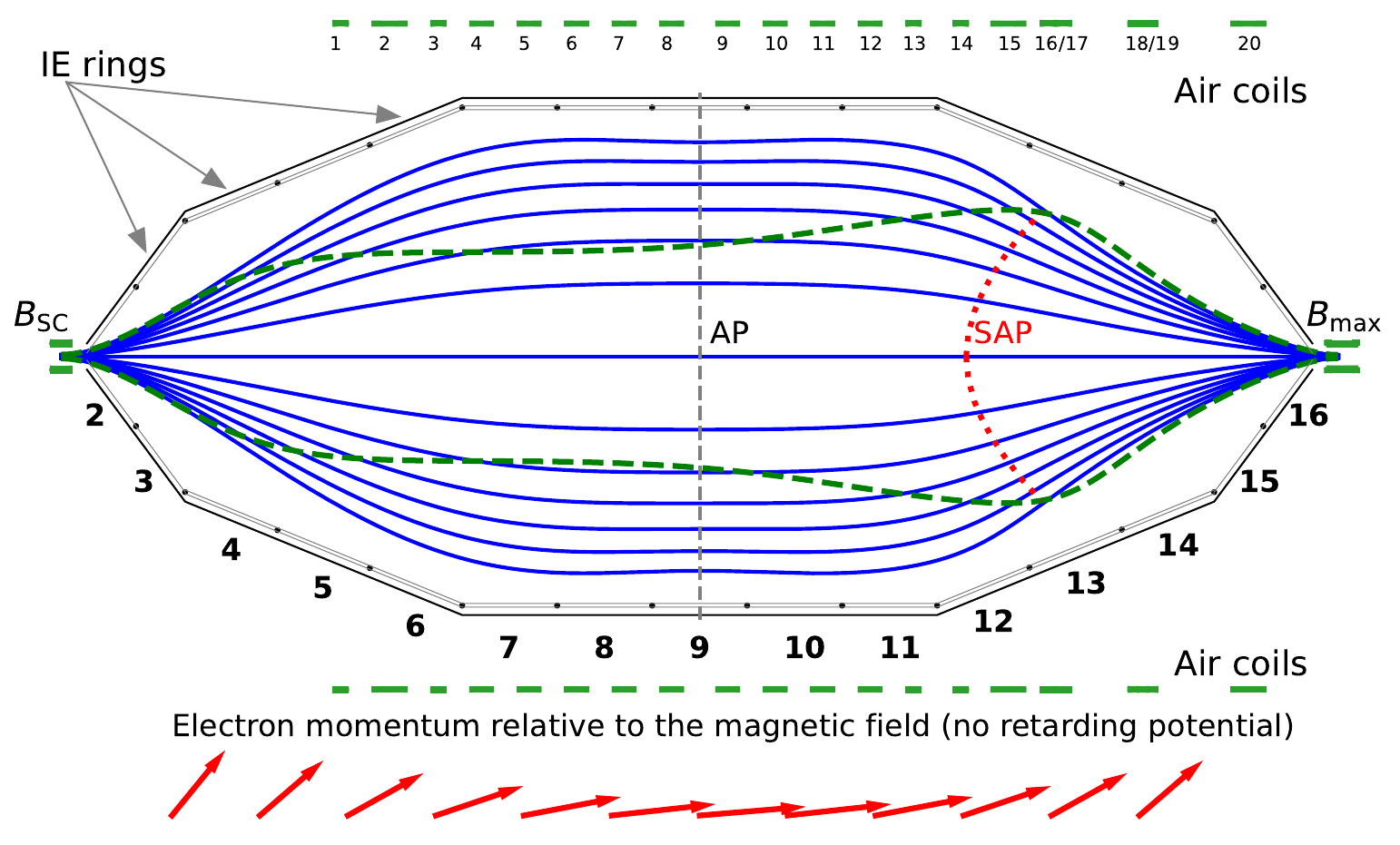}}
\caption{The MAC-E-filter principle of the KATRIN main spectrometer. The solid blue lines indicate the magnetic field lines inside the spectrometer. The magnetic field is produced by the superconducting solenoids ($B_\mathrm{SC}$ and $B_\mathrm{max}$) and 20 air coils. The electric retarding potential is generated by the voltage applied to the spectrometer vessel (solid dark gray line) and the inner electrode (IE) rings 2-16. The analysing plane for a symmetric field configuration is shown via a dashed gray line in the middle of the spectrometer. The shifted-analysing-plane position is shown via dotted red line, with the green dashed lines indicating the corresponding magnetic flux envelope.  The schematic electron-momentum transformation for a symmetric configuration is shown by the red arrows; it assumes no retarding potential applied to the spectrometer.}
\label{fig:1}       
\end{figure*}

Electrons are emitted isotropically inside the source with a high magnetic field of \SI{2.5}{\tesla}. The electrons follow the magnetic field lines along the beamline and propagate adiabatically to the main spectrometer. In the main spectrometer the magnetic field $B$ is reduced by four orders of magnitude, see Fig~\ref{fig:1}. Adiabatic motion conserves the orbital magnetic moment of electrons: $\mu \sim \frac{E_\perp}{B} = \mathrm{const}$ in a non-relativistic case ($\mu \sim \frac{p_\perp^2}{B} = \mathrm{const}$ in a relativistic case). Therefore, the component of the electron momentum $p_\perp$ that is transverse to the magnetic field $\overrightarrow{B}$ is reduced by a factor of $\sqrt{10^4}$ as well.
Without the retarding potential, the total kinetic energy $E = E_\perp + E_\parallel$ is conserved, and the momenta of electrons are aligned with the magnetic field lines, so that $E_\parallel \gg E_\perp$. To separate the electrons by their energy an electric retarding potential $qU$ is applied to the spectrometer (with $q$ being the charge of electron and $U$ the retarding voltage). The virtual surface with the minimal $E_\parallel$ is usually referred to as the \textit{analysing plane}(AP)~\cite{Gluck:2013taa}. Notice that the term ``plane'' is used for historical reasons, while in the general case the analysing plane is a curved surface. The maximal remaining energy due to the transverse motion of the electrons, $E_\perp$, defines the energy or filter width of the spectrometer transmission:

\begin{equation}
    \Delta E = 
    \frac{B_\mathrm{ana}}{B_\mathrm{max}}\frac{\gamma + 1}{2}E,
\label{eq:energy_width}
\end{equation}
where $B_\mathrm{ana}$ stands for the magnetic field in the analysing plane, $B_\mathrm{max}$ is the maximal magnetic field in the whole KATRIN beamline located in the pinch superconducting solenoid at the exit of the spectrometer (to the right in Fig.~\ref{fig:1}), and $\gamma$ is the gamma factor of the electron with kinetic energy $E$.

The probability that electrons emitted isotropically in the source with energy $E$ would reach the detector assuming the electric retarding potential $qU$ is given by a transmission function (energy losses due to scattering and synchrotron radiation are neglected here):
\begin{equation}
  T(qU, E) =
    \begin{cases}
      0 & \text{if $E - qU \leq 0$ },\\
      1-\sqrt{1-\frac{E-qU}{E}\frac{2}{\gamma+1}\frac{B_\mathrm{s}}{B_\mathrm{ana}}} & \text{if $0<E-qU \leq \Delta E$ },\\
      1-\sqrt{1-\frac{B_\mathrm{s}}{B_\mathrm{max}}} & \text{if $E-qU > \Delta E$}
    \end{cases}    
    \label{eq:transmission_isotropic}
\end{equation}
where $\Delta E$ is given by Eq.~\ref{eq:energy_width} and $B_\mathrm{s}$ is the magnetic field at the source.
The pinch magnetic field $B_\mathrm{max}$ of \SI{4.2}{\tesla} creates a magnetic mirror effect, which reflects electrons with an initial pitch angle (the angle between the momentum of the electron and the magnetic field at the source) larger than $\theta_\mathrm{max}\approx 51^\circ$. The maximal transmission probability for the electrons with a high surplus energy $E>\Delta E + qU$ is defined by the corresponding solid angle. 
With the finite size of the detector pixels and therefore the finite area of the analysing plane that is mapped onto a given detector pixel, any variation of the fields within this area would effectively smear the transmission function in Eq.~\ref{eq:transmission_isotropic}. This smearing can be approximated by an effective Gaussian broadening of the transmission function with a standard deviation $\sigma_\mathrm{ana}$.

The transmission function can be considered as the experimental response of the KATRIN main spectrometer to a monoenergetic, isotropic source of electrons. For an isotropic source with a given electron energy distribution $\frac{dN}{dt}(E)$, the measured spectrum can be described by a convolution of the differential spectrum with the transmission (or response) function $T(qU,E)$:
\begin{equation}
    R(qU) = \int T(qU, E) \frac{dN}{dt}(E) dE.
\label{eq:intSpectrumGeneral}
\end{equation}
This integrated spectrum is then scanned by applying different electric retarding potentials $qU$. 

The following corrections to Eq.~\ref{eq:transmission_isotropic} need to be considered when describing the transmission function. First, the scattering of the electrons in the source changes the angular distribution of electrons at a given energy. The electrons with higher pitch angles are more likely to scatter due to the longer paths in the source, therefore fewer non-scattered electrons with high pitch angles will arrive in the main spectrometer. Such electrons have a non-isotropic angular distribution and the isotropic transmission in Eq.~\ref{eq:transmission_isotropic} has to be modified~\cite{Kleesiek:2018mel}. The second effect is related to the energy losses of electrons due to synchrotron radiation in the strong magnetic fields. This effect changes the shape of the transmission function, smearing slightly the upper part of the transmission curve, which corresponds to high pitch angles. Third, electron scattering on the gas in the source modifies the energy and is described by the energy-loss function. The resulting experimental response is given by a convolution of the transmission function with the energy losses for 1-,2- and n-fold-scattered electrons. We will not consider the scattered electrons for the presented analysis, therefore, we do not describe here the otherwise crucial effect of energy losses due to scattering~\cite{Kleesiek:2018mel}.

The magnetic field in the main spectrometer is shaped by the superconducting solenoids at the entrance and exit of the spectrometer and is fine tuned by a set of 20 air coils around the main spectrometer, see figure~\ref{fig:1}. By varying the currents in the air coils independently, one can place the minimum of the magnetic field at various positions along the axis of the spectrometer and also change the magnitude of the field in a range of \SIrange{0}{2}{\milli\tesla}~\cite{Gluck:2013taa,Erhard:2017htg}.

The electric potential is defined by the vessel potential and additional offsets, which are applied to the inner electrode system, namely the electrode rings 2-16 in figure~\ref{fig:1}. A typical offset is about $-200$~V with respect to the vessel potential, providing an electric shielding for any low-energy electrons produced near the vessel walls~\cite{KATRIN:2021dfa,Valerius:2006qf}. 

\section{Background reduction with the shifted-analysing-plane configuration}
\label{sec:3}

The background in KATRIN has been studied extensively and several background mitigation methods were introduced~\cite{KATRIN:2018rxw,KATRIN:2018lln,KATRIN:2019dnj,KATRIN:2019mkh}.
The main background sources in KATRIN are related to radioactive decays or secondary-electron production in the main spectrometer, to the Penning trap between the main and the pre-spectrometer, and to the background events in the detector section of the KATRIN setup.
A significant part of the background electrons is generated in the volume of the main spectrometer. Such electrons are accelerated by the retarding potential towards the detector and have similar energies as the signal electrons from tritium $\upbeta$ decay.
The low-energy electrons can be produced by radioactive decay of Rn atoms emanating from the non-evaporable getter (NEG) material in the main spectrometer pump ports. Radon propagates into the volume of the spectrometer, decays there and produces high-energy primary electrons. They are likely to be trapped between the superconducting solenoids at the entrance and at the exit of the main spectrometer. They ionise the residual gas, producing low-energy secondary electrons, which are able to escape the magnetic trap and eventually reach the detector. This component of the background electrons was significantly reduced by installing liquid-nitrogen-cooled baffles between the NEG pumps and the spectrometer volume~\cite{Harms:2015zpn,Gorhardt:2018rqg}.
The background of the detector section is related to both cosmic muons hitting the detector wafer and the intrinsic noise. The cosmic-muon events are partially rejected by a dedicated muon veto system and the intrinsic noise is reduced with proper energy cuts and corresponding region-of-interest selection~\cite{Amsbaugh:2014uca}.

The main remaining contribution to the background events is caused by radioactive decays of $^{210}$Pb that was implanted into the surface of the walls of the main spectrometer vessel while it was exposed to ambient air containing radon. The decay chain ends with the $^{210}$Po $\upalpha$ decay into stable $^{206}$Pb. During $\upalpha$ decay near the surface of the stainless-steel walls of the vessel, a sputtering process can occur, 
producing highly excited atoms (e.g. hydrogen or oxygen) with one electron in an highly excited orbital (Rydberg atom) or with even two electrons in excited orbitals. These highly excited atoms can propagate into the volume of the spectrometer and can be ionised there by black-body radiation (Rydberg) or via autoionisation (double excited). The resulting electrons have low energies of a few 10~meV (Rydberg) or a few 100~meV (autoionising), and are distributed almost uniformly within the volume~\cite{KATRIN:2020zld,Trost:2019pey,Hinz:2022gaf}.

\begin{figure*}[h]
  \frame{\includegraphics[width=0.75\textwidth,center]{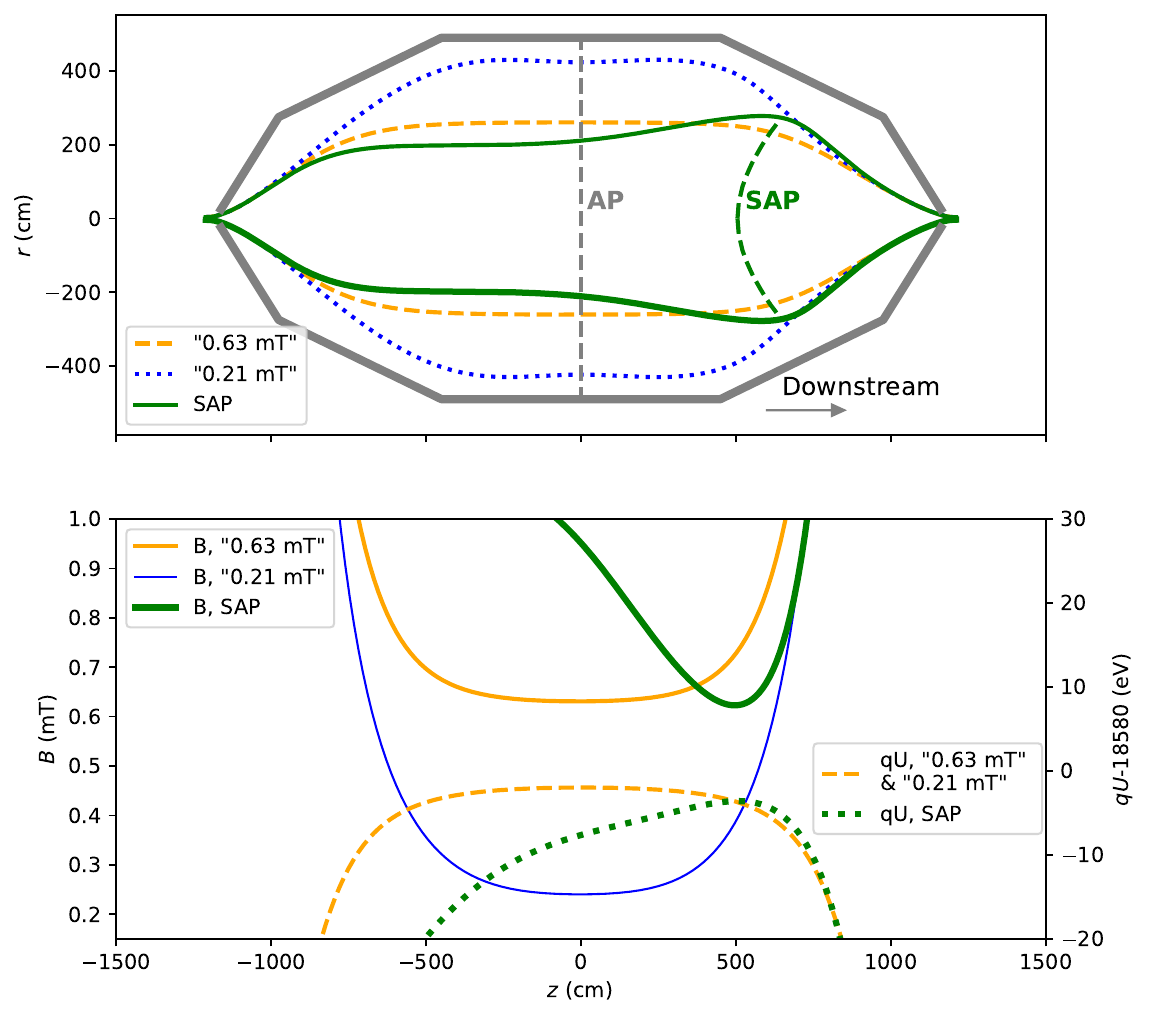}}
\caption{Top: Magnetic flux for the nominal symmetric and shifted analysing plane configurations inside the spectrometer vessel (thick gray lines). The flux envelopes of the symmetric configurations are shown with the dotted blue line ($B_\mathrm{ana}=\SI{0.21}{\milli\tesla}$) and dashed orange line ($B_\mathrm{ana}=\SI{0.63}{\milli\tesla}$). The flux for the SAP is shown by a green solid line. Bottom: The profiles of the electric retarding potential and the magnetic field along the axis of the spectrometer. The figure is adapted from \cite{Lokhov:2022iag}.}
\label{fig:2}       
\end{figure*}

The MAC-E-filter principle of the main spectrometer requires a small magnetic field in the analysing plane to reach an energy resolution on the order of \SI{1}{\electronvolt}. The radius of the magnetic flux is thus increased by a factor of 100, from \SI{10}{\centi\meter} in the source to up to \SI{10}{\meter} inside the spectrometer.
When produced in the region of the highest retarding potential, the electrons are accelerated towards the detector (downstream) or towards the source section (upstream). Only the downstream part of the magnetic flux contributes to the background counts at the detector, see Fig.~\ref{fig:2}. To reduce the background rate in the first two neutrino-mass campaigns the magnetic field in the analysing plane was increased from the design value of \SI{0.21}{\milli\tesla} to \SI{0.63}{\milli\tesla}. The downstream volume was thereby reduced from \SI{354}{\meter^3} to \SI{160}{\meter^3}. However, the remaining background rate was around 220 to 300~mcps, still exceeding the design expectation by more than an order of magnitude~\cite{KATRIN:2021dfa}.
To further exploit the volume-dependency of the background rate one can shift the analysing plane towards the detector to reduce the downstream volume that is mapped onto the detector. This is the concept of the shifted-analysing-plane configuration of the electromagnetic fields in the KATRIN main spectrometer~\cite{Dyba:2019mdq,Schaller:2020kem,Lokhov:2022iag}.

The idea of the shifted-analysing-plane configuration is to move the region of the highest retarding potential by approximately \SI{6}{\meter} towards the detector, significantly reducing the downstream volume of the flux tube.
The main challenge of such a configuration is to place the minima of the magnetic field and the maxima of the electric potential at the same positions. A suitable combination of the electric and magnetic field setting was found via simulation as described in Ref.~\cite{Lokhov:2022iag}. 

\begin{figure*}[h]
  \frame{
  \includegraphics[width=0.45\textwidth]  {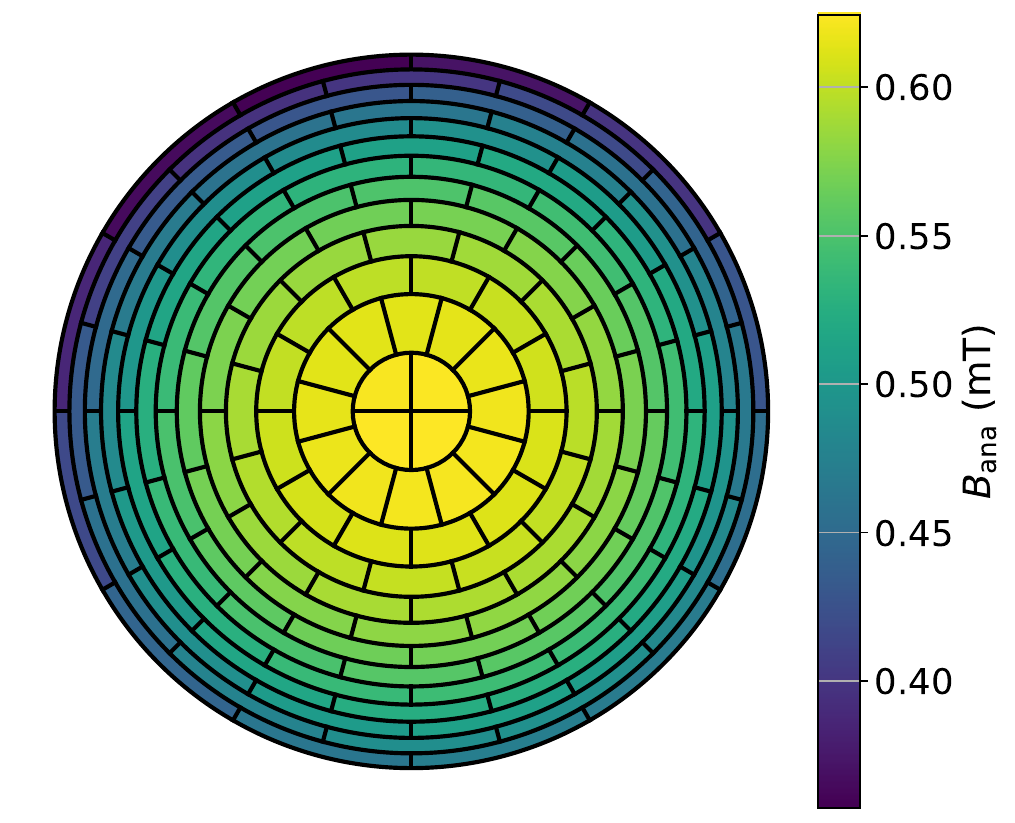}
  \includegraphics[width=0.45\textwidth]{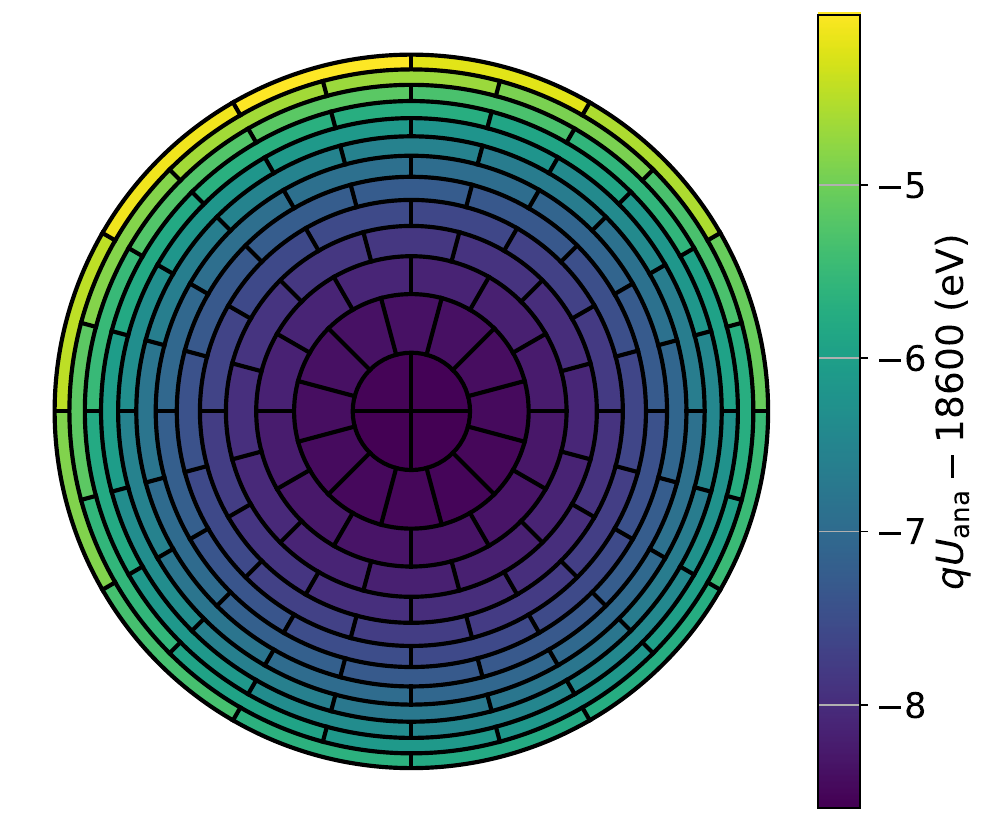}
  }
\caption{ An image of the magnetic field (left) and the electric retarding potential (right) of the SAP configuration on the detector pixels. The magnetic field in the analysing plane is estimated for the electrons arriving at the center of each pixel via a full simulation of electromagnetic fields in the main spectrometer. A typical variation of \SI{0.2}{\milli\tesla} for the SAP mode is significantly larger than the magnetic-field variation in a symmetric configuration ($\approx\SI{0.004}{\milli\tesla}$). The variation of retarding potential $qU_\mathrm{ana}$ is on the order of $\mathcal{O}(\SI{1}{\electronvolt})$ in the SAP mode, while the variation in a symmetric setting only reaches $\mathcal{O}(\SI{0.14}{\electronvolt})$.  A slight misalignment of the system is visible in the axially asymmetric distribution of fields in the outer pixels of the detector in both figures.}
\label{fig:3}       
\end{figure*}

Since 2020 the SAP configuration of the electromagnetic fields in the main spectrometer is the new default operation mode in KATRIN.
The demonstrated reduction of the spectrometer background rate in the SAP configuration reaches approximately \SI{50}{\percent}. 
However, the layout of the electromagnetic fields in the spectrometer becomes more complex, compared to the nominal symmetrical configuration, see figure~\ref{fig:2}. Moving the analysing plane from the central, cylinder-shaped part of the vessel to the cone-shaped part results in a significantly higher inhomogeneity of the fields: magnetic-field variations of \SI{0.2}{\milli\tesla} and electric-potential variations of about \SI{3}{\volt} within the analysing plane, see figure~\ref{fig:3}. The geometry of the setup defines the magnetic field and the electric potential at a given point in the analysing plane. The fields cannot be obtained from simulations with a sufficient precision because of imperfect knowledge of the as-built alignment. Therefore, the transmission properties of the spectrometer have to be measured independently using an established calibration source.

\section{Electromagnetic field measurement with $^\mathrm{83m}$Kr}
\label{sec:4}
Gaseous $^\mathrm{83m}$Kr is used by the KATRIN experiment as a well established calibration source of electrons. $^\mathrm{83m}$Kr is generated in the setup through the decay of $^{83}$Rb with a total activity up to \SI{10}{\giga\becquerel}. It emanates from a zeolite source with an efficiency up to \SI{90}{\percent} and a negligible release of $^{83}$Rb~\cite{Venos:2014pra,VENOS2005323,Sentkerestiova:2018oth}.

$^\mathrm{83m}$Kr gas can be either used in the so-called pure Kr mode or co-circulated with the tritium (or other carrier gas) in the source. The advantage of the latter mode is the possibility to achieve high rates of $^\mathrm{83m}$Kr decay in the source. 
The fluctuations or drift of the gas pressure in the source in this mode are below \SI{1}{\percent} on the time scale of several hours~\cite{Marsteller:2020loy,KATRIN:2022zqa}. The amount of $^\mathrm{83m}$Kr atoms in the source corresponds to less than 1~ppm of the tritium gas and does not modify the gas properties.

$^\mathrm{83m}$Kr undergoes a cascade of two gamma transitions with energies of \SI{32.2}{\kilo\electronvolt} and \SI{9.4}{\kilo\electronvolt}. Both transitions are highly converted and provide electrons with high intensity~\cite{Venos:2018tnw}.

For the calibration of the KATRIN setup, mainly the conversion electrons of $^\mathrm{83m}$Kr from the subshells K, L$_3$, N$_2$ and N$_3$ of the \SI{32.2}{\kilo\electronvolt} transition are used. The corresponding lines have energies of \SIrange{17.8}{32.2}{\kilo\electronvolt} and narrow natural line widths of less than \SI{3}{\electronvolt}.
The K-32 line has an the energy of \SI{17.8}{\kilo\electronvolt}, about \SI{0.8}{\kilo\electronvolt} below the tritium endpoint, and a natural line width of about \SI{2.7}{\electronvolt}, making it an excellent calibration tool for the neutrino-mass measurements.
The L$_3$-32 line has an energy of \SI{30.5}{\kilo\electronvolt}, a line width of about \SI{1}{\electronvolt} and relatively high intensity. It can be used as a tool for quick measurements.
Finally, the N$_{2,3}$-32 doublet lines with energies of about \SI{32.1}{\kilo\electronvolt} have negligibly narrow natural line widths, well below the energy resolution of the spectrometer. Even though the intensity of these lines is comparatively low, the vanishing line widths make this doublet a high-precision tool for measuring fine effects, e.g. a spectrum broadening of less than \SI{50}{\milli\electronvolt}.

The parameters of the lines are known with good accuracy and were measured extensively~\cite{Venos:2018tnw}, including a high-precision measurement in the KATRIN setup~\cite{Altenmuller:2019ddl}.
However, to fulfill the precision requirements of the neutrino-mass determination, the parameters of the transmission function have to be known on the sub-percent level. Any unaccounted broadening $\sigma$ of the spectrum leads to a systematic bias of the squared neutrino mass by $\Delta m_\nu^2 = - 2\sigma^2$~\cite{Robertson:1988xca}. A bias of the magnetic field in the analysing plane of \SI{1}{\percent} can cause a bias of the order of \SI{0.01}{\electronvolt^2}.
In order to measure the transmission broadening and the magnetic field in the analysing plane with the $^\mathrm{83m}$Kr conversion electrons directly, the differential spectrum of the corresponding lines has to be known with a sub-percent precision as well.

To minimise systematic uncertainties of the $^\mathrm{83m}$Kr spectra in the spectrometer-field estimation, we use the method of reference measurement. The $^\mathrm{83m}$Kr lines spectra are scanned using two different configurations of the main spectrometer fields: the symmetric (nominal) one with the magnetic field $B_\mathrm{ana}\sim \SI{0.27}{\milli\tesla}$ and the SAP configuration. The measurements are performed back-to-back so that there is no change in any other conditions except for the fields in the main spectrometer. Since the symmetric configuration can be simulated precisely and does not suffer from the minor misalignment of the setup, this measurement is used as a reference to determine the effective parameters (line energy and width) of the $^\mathrm{83m}$Kr spectrum. The parameters are considered to be only ``effective'' because of the influence of source effects (starting potential distribution, scattering, etc.)~\cite{Machatschek2021_1000132391}. These additional effects equally modify the measured spectrum in both symmetric and SAP field configuration. Therefore, obtaining the ``effective'' parameters of the $^\mathrm{83m}$Kr lines spectra in the symmetric configuration measurement and using them in the analysis of the spectra measured in the SAP mode allows extraction of the SAP magnetic field $B_\mathrm{ana}$, electric potential $qU_\mathrm{ana}$ and the transmission broadening $\sigma_\mathrm{ana}$. The effects that are not related to the electromagnetic field in the main spectrometer would be cancelled out by this technique.

\section{Analysis of the K-32 line spectrum}
\label{sec:5}
\begin{figure*}[h]
  \frame{\includegraphics[width=0.95\textwidth,center]{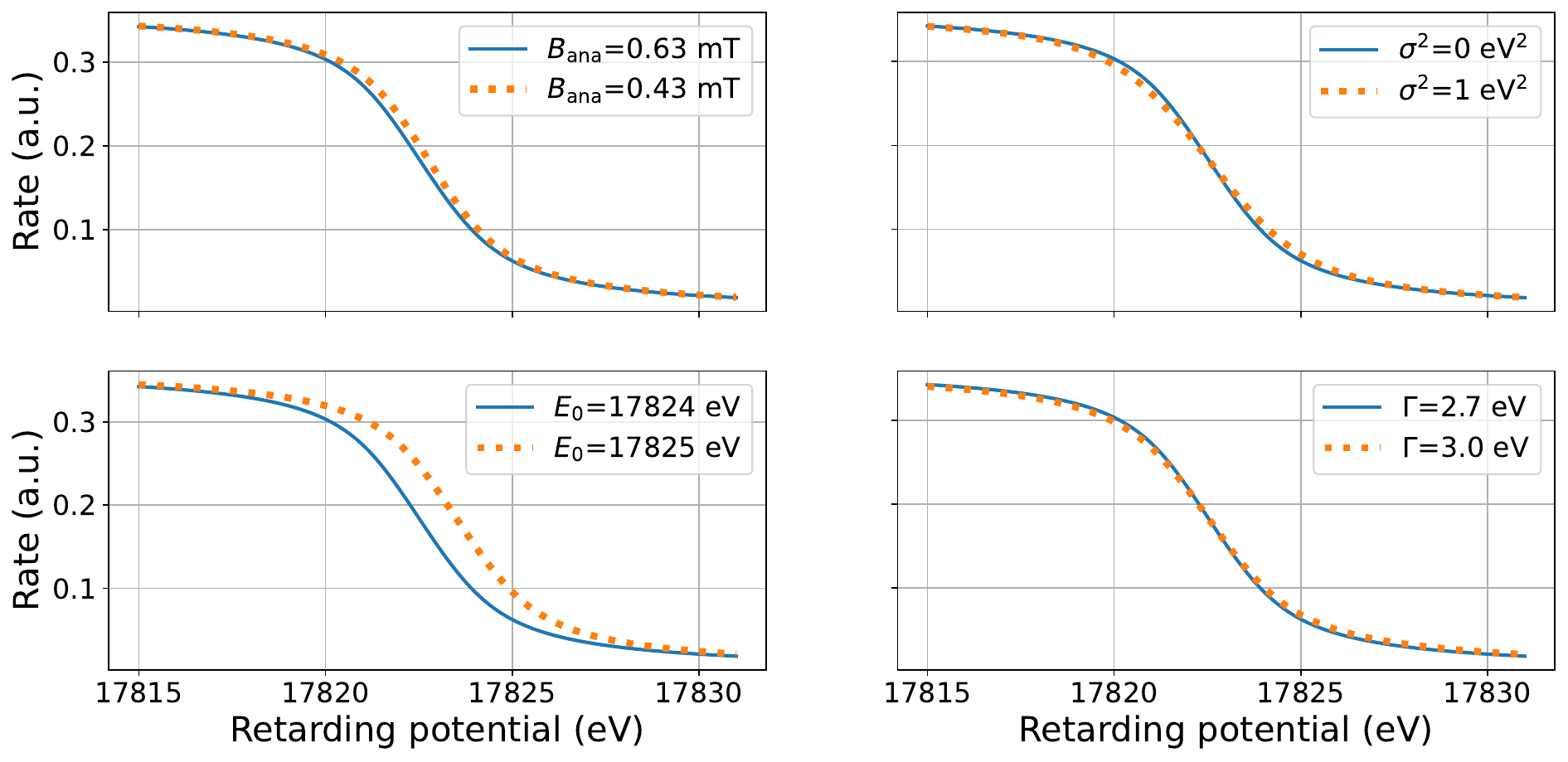}}
\caption{Schematic illustration the integrated spectrum of the K-32 line. The solid blue line shows the spectrum for $E_0=\SI{17824}{\electronvolt}$, $B_\mathrm{ana}=\SI{0.63}{\milli\tesla}$, $\Gamma =\SI{2.7}{\electronvolt}$ and broadening $\sigma^2 = \SI{0}{\electronvolt^2}$. In each of the sub-figures, the orange dotted line shows the spectrum for one of these four parameters being modified. Notice that the size of the parameter change is exaggerated to visualize the impact of the corresponding parameter on the spectrum.}
\label{fig:4}       
\end{figure*}

The integrated spectrum measured by KATRIN is modelled as a convolution of the transmission function with the differential spectrum of the corresponding $^\mathrm{83m}$Kr line, see Eq.~\ref{eq:intSpectrumGeneral}.
The energy spectrum of the K-32 line electrons is described by a Lorentzian with the natural line width $\Gamma$, line position $E_0$ and normalization factor $A$~\cite{Altenmuller:2019ddl}:

\begin{equation}
    L(E;A,E_0,\Gamma)=\frac{A}{\pi}\frac{\Gamma/2}{(E-E_0)^2+\Gamma^2/4}.
\label{eq:loretzian}
\end{equation}
The broadening effects (e.g. Doppler thermal broadening) are taken into account by convolving the Lorentzian with a Gaussian function, resulting in a so-called Voigt function $V$:

\begin{equation}
    V(E;A,E_0,\Gamma,\sigma) = L(E;A,E_0,\Gamma) \otimes \left( \frac{1}{\sqrt{2\pi\sigma^2}} e^{-\frac{E^2}{2\sigma^2}} \right).
\label{eq:voigt}
\end{equation}
The measured integrated spectrum shape is given by Eq.~\ref{eq:intSpectrumGeneral} with $\frac{dN}{dt}(E) = V(E)$. An illustration of the integrated spectrum of the K-32 line is shown in figure~\ref{fig:4}.

Since the mean energy of the K-32 electrons is lower than the endpoint energy of tritium, $E_0=\SI{18.6}{\kilo\electronvolt}$, a significant contribution to the background rate stems from the $\upbeta$ decay of the residual tritium gas in the source and tritium accumulated on the rear wall upstream of the WGTS. The tritium spectrum is approximated as an additional background rate, linearly depending on the retarding potential $qU$ with a coefficient $a_\mathrm{slope}$; the linear approximation is valid due to a narrow scan range of \SI{30}{\electronvolt} compared to the K-32 line distance to the endpoint, $E_0-E_\mathrm{K-32}\sim \SI{800}{\electronvolt}$. $^\mathrm{83m}$Kr conversion electrons from lines at higher energies create an additional background rate $R_\mathrm{bg}$, which is added to the integrated spectrum as well.

The measurement of the K-32 line was performed in pure Kr mode and contained 32 spectral scans in the symmetric nominal configuration (\SI{12}{\hour} of live time, 36 scan steps in the range $qU=\SIrange{17819}{17850}{\electronvolt}$) and 91 spectral scans in the SAP configuration (\SI{35}{\hour} of live time, 36 scan steps in the range $qU=\SIrange{17824}{17855}{\electronvolt}$). 
For each mode the acquired scans are combined by adding the counts and the live times for the same $qU$. The combination is possible due to the high stability and reproducibility of the retarding potential~\cite{Rodenbeck:2022iys}. The narrow analysis range allows us to neglect the scattering of electrons on the residual gas in the source: due to inelastic scattering, electrons with a starting energy close to the line position $E_0$ lose more than \SI{10}{\electronvolt} of kinetic energy and are effectively removed from the analysis window.

\textbf{\textit{Pixel combination.}}
Each of the 148 pixels of the focal-plane detector of KATRIN records an independent spectrum of the $^\mathrm{83m}$Kr line. 
Each spectrum has its own model with individual parameters for the transmission function, background and detection efficiency (we will refer to such model parameters as parameters of a given pixel). The analysis of such segmented data becomes computationally expensive and numerically less stable, due to smaller statistics at each data point. To facilitate the analysis 
it is preferable to combine the data from several pixels into a single spectrum and to have one model describing it. In the first neutrino-mass measurements it was possible to combine the data from all pixels into a single spectrum using the homogeneity of the transmission function over the analysing plane in the nominal symmetric configuration (so-called uniform analysis). With the inhomogeneous electromagnetic fields in the SAP configuration, the combined pixel groups must be smaller. This avoids excessive broadening of the uniform transmission function ($\Delta qU\sim\SI{3}{\electronvolt}$), which would have to be measured with a sub permille precision. Instead one selects the pixels with similar transmission parameters and combines them into a \textit{patch} of pixels. 

In an ideally aligned system the patches would coincide with the concentric rings of the detector pixels due to the axial symmetry of the setup. However, due to a small geometrical misalignment of the setup (on the order of \SI{1}{\milli\meter} w.r.t. the \SI{10}{\centi\meter} diameter of the magnetic flux at the detector), the pixels in one detector ring can have significantly different transmission parameters. For selecting a suitable set of patches the following procedure is used.

\begin{figure*}[h]
  \frame{\includegraphics[width=0.95\textwidth,center]{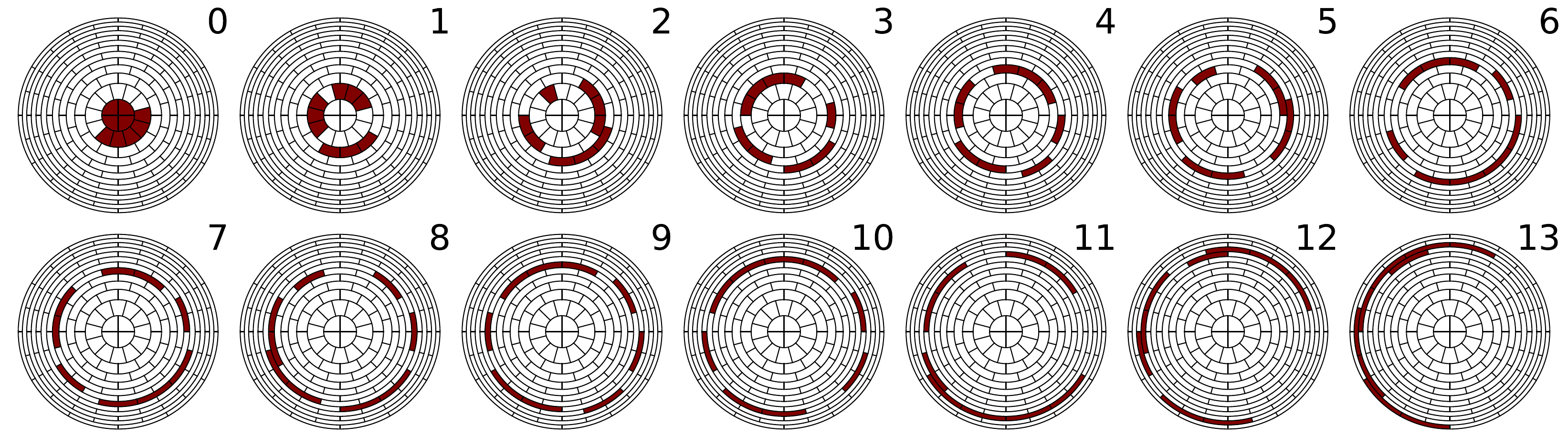}}
\caption{Grouping of the detector pixels with similar transmission parameters into patches. The patches follow general ring structure but have obvious shifts w.r.t. the detector rings. 22 pixels in the outer rings are shadowed by the beamtube and are excluded from the analysis.}
\label{fig:5}       
\end{figure*}

First, the SAP-mode-scan data for the K-32 line is analysed individually for each pixel. To make the fitting more stable, the natural line width $\Gamma$ and the broadening $\sigma$ are fixed parameters, and only the normalization factor $A$, the magnetic field $B_\mathrm{ana}$ and the line position $E_0$ are fitted (the line position being a proxy for the retarding potential $qU_\mathrm{ana}$). 126 pixels are selected for which the corresponding normalization factor is above a threshold of $0.9 A_\mathrm{max}$; the remaining 22 pixels have a lower rate due to shadowing of magnetic flux in the beamline. The selected pixels are grouped into 14 patches of 9 pixels each, according to the groupings of middle-of-transmission position (the energy $E$, where the transmission $T(qU,E) = 1/2$). This quantity is used instead of $qU_\mathrm{ana}$ and $B_\mathrm{ana}$ as a more general, symmetric characteristic of the transmission. The variation of the middle-of-transmission between pixels is mainly defined by the $qU_\mathrm{ana}$ of each pixel, with an additional correction from the changing magnetic field, see Fig.~\ref{fig:3}. The defined patches are shown in figure~\ref{fig:5}.

\begin{figure*}[h]
  \frame{\includegraphics[width=0.95\textwidth,center]{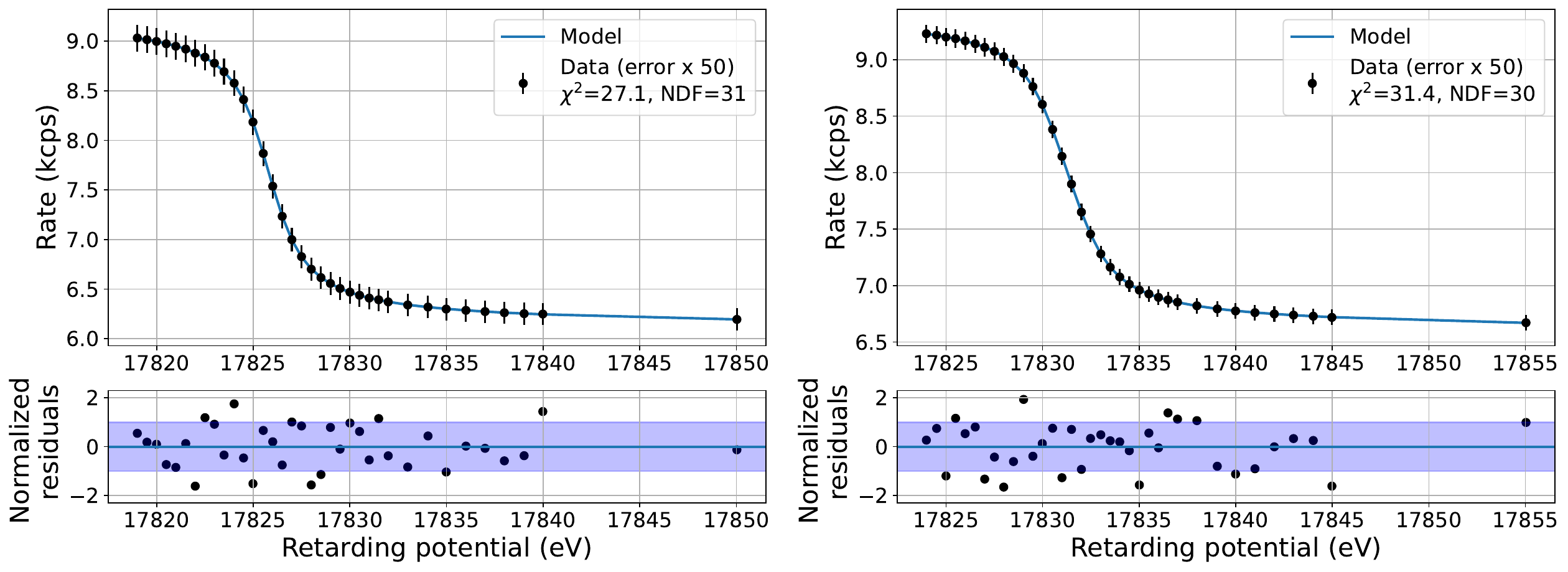}}
\caption{An example of measured spectra in patch 0 of the K-32 line of $^\mathrm{83m}$Kr in the symmetric analysing plane (left) and shifted analysing plane (right) configurations with the fit and the residuals. The effective line position differs for the two configurations due to the different potential depression. The simple goodness-of-fit parameter $\frac{\upchi^2}{N_\mathrm{DoF}}$ is typically close to 1. The uncertainties are scaled by a factor of 50 to be visible in the plot.}
\label{fig:6}       
\end{figure*}

\textbf{\textit{Analysis of the reference measurement in symmetric configuration.}}
The parameters of the K-32 line spectrum are inferred via the method of maximum likelihood assuming a Gaussian distribution of number of counts at each scan step; in this case $- 2 \ln{\mathcal{L}}$ corresponds to the $\chi^2$ function (Eq.~\ref{eq:likelihood}). For each patch the negative logarithm of the likelihood function is minimized w.r.t. the fit parameters:
\begin{equation}
    - 2 \ln{\mathcal{L}}(A,E_0,\Gamma^2,B_\mathrm{ana},R_\mathrm{bg},a_\mathrm{slope}) = \sum_i \frac{(N_i-\mu_{i,\mathrm{model}}\cdot T_i)^2}{\mu_{i,\mathrm{model}}\cdot T_i}+\mathrm{pull\ terms},
\label{eq:likelihood}
\end{equation}
where $N_i$ is the number of counts in the scan step $i$ in the respective patch, $T_i$ is the measurement time at the retarding potential $qU_i$ and $\mu_{i,\mathrm{model}}$ is the model prediction of the rate, $R_\mathrm{bg}$ and $a_\mathrm{slope}$ are the constant and linear background rate parameters. The summation goes over all the scan steps $i$ in the analysis window.
The Gaussian broadening of the line is fixed to the value of the thermal Doppler broadening of \SI{0.041}{\electronvolt} for the energy of \SI{17.8}{\kilo\electronvolt} and the gas temperature of \SI{80}{K}. The magnetic field in the analysing plane $B_\mathrm{ana}$ is constrained by a pull term $-2 \ln{\mathcal{L}_\mathrm{pull}} = \frac{(B_\mathrm{ana}-B_\mathrm{sim})^2}{(\SI{5}{\micro\tesla})^2}$. $B_\mathrm{sim}$ is taken from the simulations of the symmetric analysing plane configuration with KASSIOPEIA software~\cite{Furse:2016fch}. The uncertainty of \SI{5}{\micro\tesla} is derived from the discrepancy of the simulation results and dedicated measurements with magnetometers~\cite{Letnev:2018fkq,Block2022_1000145073}. An example of the fit is shown in figure~\ref{fig:6}.

The parameters of interest of the reference measurement are the effective line width $\Gamma^2_\mathrm{i,sym}$ and the line position $E_\mathrm{0,\,i,sym}$ for each patch $i$, which will be used as inputs for fitting of the SAP scans of the K-32 line.

\textbf{\textit{Analysis of the K-32 line scans in the SAP configuration.}}
The fit of the K-32 line spectrum in the SAP mode for each patch is performed using the same likelihood, Eq.~\ref{eq:likelihood}, where the pull term for $B_\mathrm{ana}$ is replaced with the one for the squared line width $\Gamma^2$: $-2 \ln{\mathcal{L}_\mathrm{pull}} = \frac{(\Gamma^2-\Gamma^2_\mathrm{i,sym})^2}{\sigma_{\Gamma^2_\mathrm{i,sym}}^2}$. This line width includes any possible broadening which comes from the source potential in both measurement modes (nominal or symmetric and SAP).
As opposed to the reference measurement, both the magnetic field $B_\mathrm{ana}$ and the broadening $\sigma^2$ are the free parameters of this fit.

\begin{figure*}[h]
  \frame{\includegraphics[width=0.75\textwidth,center]{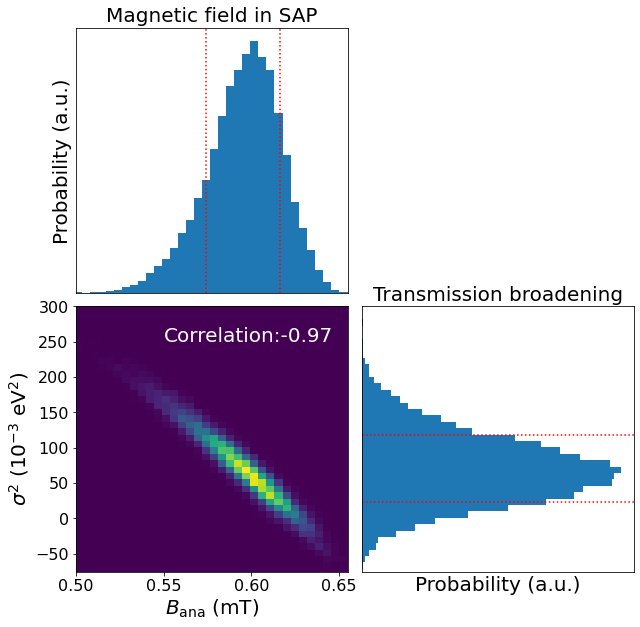}}
\caption{An example of the Markov Chain Monte Carlo estimation of the uncertainties and the correlation between the magnetic field and the transmission broadening for patch 0. The ellipse of the 2-d distribution indicates the high anticorrelation of the two parameters. The red dotted lines show the 1-$\sigma$ uncertainty interval.}
\label{fig:7}       
\end{figure*}

One challenge of the parameter inference and the error estimation is the high anticorrelation of $B_\mathrm{ana}$ and $\sigma^2$. The two parameters have a similar impact on the shape of the spectrum: an increase of the effective line width can be described as either a higher magnetic field or a higher transmission broadening. To make a numerically stable estimation, a gradient-based minimization is used to determine the minimum of $-2 \ln{\mathcal{L}}$, see figure~\ref{fig:6}. Then, a Markov Chain Monte Carlo (MCMC) method is used to profile the likelihood around the minimum and derive the uncertainties and correlation of the two parameter; see the example of such an MCMC sample in figure~\ref{fig:7}. An independent analysis was performed using a grid scan of the likelihood near the minimum and yielded the same uncertainty estimation. Notice that the relatively large uncertainty of the $\sigma^2$ parameter allows for fluctuations of $\sigma^2$ into unphysical negative values. To avoid setting limits on this parameter and to preserve the symmetry of the likelihood function, a mathematical continuation to the negative $\sigma^2$ values is used following the prescription in~\cite{Belesev:2008zz}.

The final step of determining the transmission function in the SAP mode is the determination of the retarding potential $qU_\mathrm{ana}$ for a given patch, when the retarding potential $qU$ is applied to the spectrometer. The difference between the two values is referred to as the ``potential depression'' and it is related to the different potentials applied to the spectrometer walls and the inner electrode system.
The inferred effective line position of the reference measurement is corrected for the simulated potential depression in the symmetric field configuration. Then the corrected line position is compared to the effective line position fitted in the SAP measurement. Since the physical line position is independent from the electromagnetic field configuration, the difference of the line positions yields the potential depression of the SAP mode.
\begin{align}
    E_{0,\mathrm{sym,fit}}=E_{0,\mathrm{true}}+(qU-qU_\mathrm{sym,ana}); \label{eq:depression1} \\
    E_{0,\mathrm{SAP,fit}}=E_{0,\mathrm{true}}+(qU-qU_\mathrm{SAP,ana}).
\label{eq:depression2}
\end{align}
Subtracting Eqs.~\ref{eq:depression1},\ref{eq:depression2}, the line position $E_{0,\mathrm{true}}$ and the potential $qU$ cancel out, and the potential in the shifted analysing plane $qU_\mathrm{SAP,ana}$ is recovered as a function of the measured line positions and the simulated potential of the symmetric setting.

The potential depression between $qU_\textrm{K-32}=\SI{17.8}{\kilo\electronvolt}$ and $qU_\mathrm{endpoint}=\SI{18.6}{\kilo\electronvolt}$ can vary for all the patches in SAP mode due to an additional impact of the ground electrode at each end of the spectrometer ($qU_\mathrm{ground}=\mathrm{const}=\SI{0}{\electronvolt}$), but the change is smaller than \SI{10}{\milli\electronvolt}. The effect is even smaller for the symmetric configuration, because the distance of the analysing plane to the ground electrode is larger. The impact of the ground electrode is more pronounced for the N$_{2,3}$-32 measurement because of the higher energy difference to the tritium endpoint.

The resulting values for $qU_\mathrm{ana}$, $B_\mathrm{ana}$, $\sigma^2_\mathrm{ana}$ and the correlation $\rho$ between the magnetic field and the broadening for each patch are shown in figure~\ref{fig:11}.

The uncertainties are dominated by the statistics of the measurement.
To investigate possible systematic effects of the other parameters of the model the following procedure was implemented. The two magnetic fields $B_\mathrm{max}$ and $B_\mathrm{s}$, which enter the transmission function in Eq.~\ref{eq:transmission_isotropic}, are added as fit parameters, constrained via pull terms. A negligible impact on the uncertainties and small correlations (below 0.1) to the parameters of interest were found.

The measurement of the electromagnetic fields in the shifted analysing plane with the K-32 line of $^\mathrm{83m}$Kr demonstrated the principle and reached the sufficient precision. The determined transmission parameters were used for analysis of the neutrino-mass measurements performed in 2020 in the SAP mode~\cite{Aker:2024drp}.

\section{Analysis of the N2,3-32 doublet measurement}
\label{sec:6}

After the proof-of-principle calibration measurement with the K-32 line, a further improvement was achieved by using the N$_{2,3}$-32 line doublet for a high-precision determination of the spectrometer fields with a pixel-wise spatial resolution. The negligible natural line widths make the lines an ideal monoenergetic source of electrons sensitive to a small experimental broadening on the order of $\mathcal{O}(\SI{0.1}{\electronvolt})$.

\begin{figure*}[h]
  \frame{\includegraphics[width=0.95\textwidth,center]{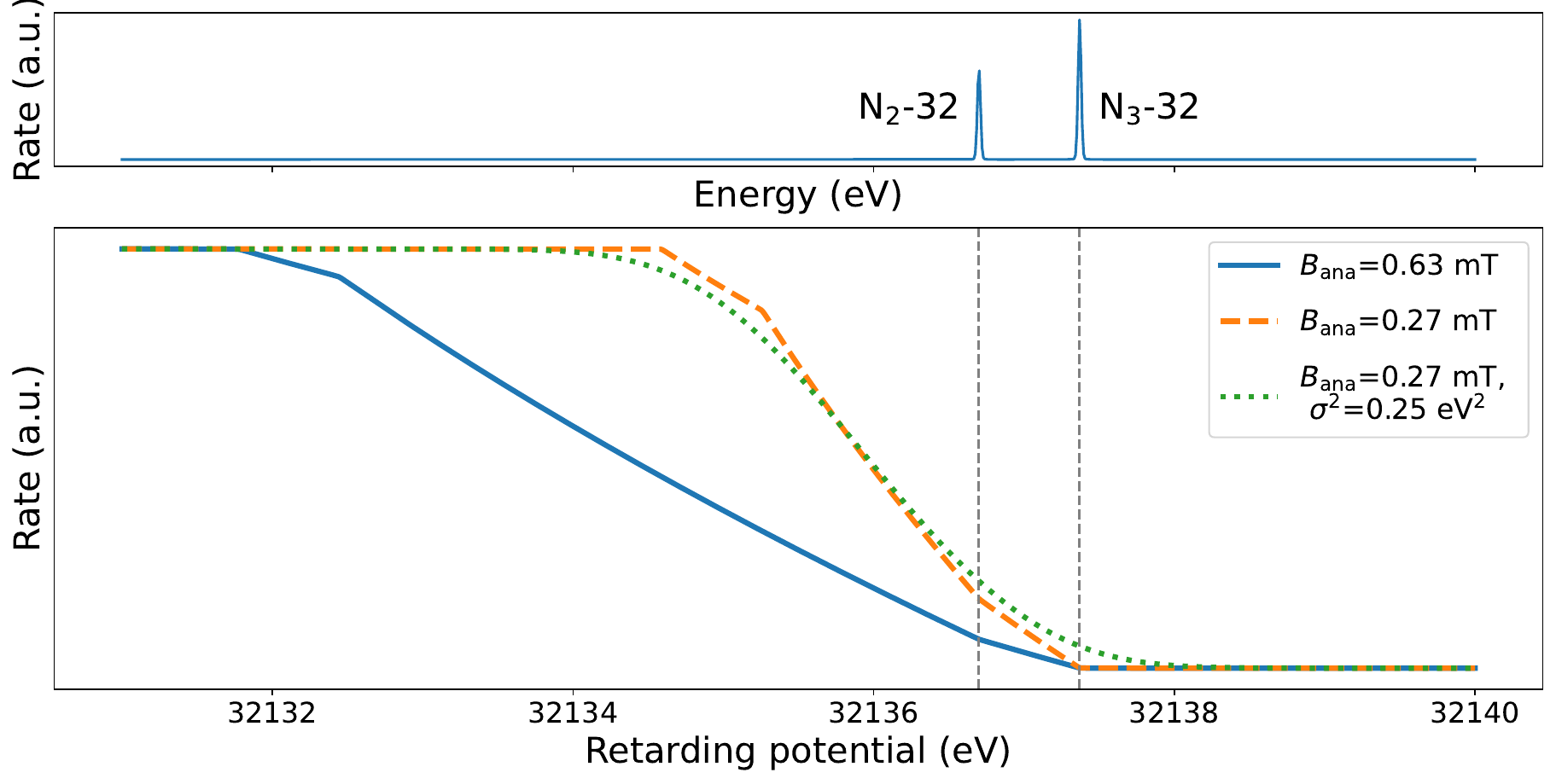}}
\caption{Top: Differential spectrum of the N$_{2,3}$-32 lines of $^\mathrm{83m}$Kr. The two lines have negligible natural line width and their separation is about \SI{0.67}{\electronvolt}. The energy of the N$_2$-32 line is set to $E_0=\SI{32136.7}{\electronvolt}$ and the relative intensity of N$_2$-32 line is $1.5$ times lower than that of N$_3$-32~\cite{Venos:2018tnw}.
Bottom: The integrated spectrum of N$_{2,3}$-32 lines for three different set of transmission parameters. The solid blue line shows the spectrum for a magnetic field $B_\mathrm{ana}=\SI{0.63}{\milli\tesla}$. The other two lines show the spectra for $B_\mathrm{ana}=\SI{0.27}{\milli\tesla}$ and the transmission broadening of $\sigma^2=\SI{0}{\electronvolt^2}$ (orange dashed) and $\sigma^2=\SI{0.25}{\electronvolt^2}$ (green dotted).}
\label{fig:8}       
\end{figure*}

The differential spectrum rate is described by two Gaussian functions (the limit of the Voigt function for $\Gamma=0$) with a common broadening $\sigma$ at a separation of about $\SI{0.67}{\electronvolt}$ with an energy of \SI{32.14}{\kilo\electronvolt}, see the top panel of figure~\ref{fig:8}. The integrated spectrum is given by a convolution of the two Gaussian functions with the transmission function in Eq.~\ref{eq:transmission_isotropic}. Three examples of modelled integrated spectra for different $B_\mathrm{ana}$ and $\sigma$ values are shown in the lower part of figure~\ref{fig:8}.

The lower intensity of the N$_{2,3}$-32 doublet compared to the K-32 and L-32 lines requires either longer measurement times or higher $^\mathrm{83m}$Kr activity in the source.
For this measurement the source configuration with co-circulation of krypton and tritium was chosen. Tritium works as a carrier gas and delivers a high amount of $^\mathrm{83m}$Kr atoms into the source. The tritium column density in this mode reaches about \SI{40}{\percent} of the nominal value (\SI{75}{\percent} density is typically used for the neutrino-mass measurements). The high-density tritium gas creates a plasma in the source, shielding the work-function inhomogeneities of the source walls and therefore reducing the overall experimental broadening, compared to the pure Kr mode. Notice that the co-circulation of tritium and krypton would not allow the measurement of the K-32 line, since the rate of $\upbeta$-decay electrons would overwhelm $^\mathrm{83m}$Kr electron spectrum at the energy of \SI{17.8}{\kilo\electronvolt}.

For the precise description of the $N_{2,3}$-32 lines, two additional effects have to be included in the model of the transmission function. First, the presence of tritium in the source increases the probability of electron inelastic scattering. As mentioned in Sec.~\ref{sec:2}, the isotropic transmission function in Eq.~\ref{eq:transmission_isotropic} has to be modified to account for the resulting anisotropic distribution of non-scattered electrons~\cite{Kleesiek:2018mel}. The column density, which affects the angular distribution, is considered as a constrained nuisance parameter with a conservative uncertainty of \SI{5}{\percent}. As in the case of the K-32 line measurements, the analysis window is chosen to exclude all scattered electrons, which lose at least $\mathcal{O}(\SI{10}{\electronvolt})$ of kinetic energy per scatter.
The second effect is related to the electron detection efficiency, which depends on the angle with which the electrons impinge on the detector. The \SI{0}{\degree} angle implies the highest detection probability, while at larger angles the probability of electrons to scatter back at the detector and to fall out of the region of interest cut increases. To correct for this effect a simple second order polynomial is introduced in the derivation of the transmission function (Eq.~\ref{eq:transmission_isotropic}) to describe the relative detection efficiency for different incident angles of electrons.

The data set consists of 16 scans performed in the symmetric field configuration (\SI{12.5}{\hour} total measurement time, 41 scan steps in the $qU$-range of \SIrange{32135}{32142}{\electronvolt}) and 21 spectrum scans in SAP mode (\SI{16.5}{\hour} measurement time, 41 scan steps in the $qU$-range of \SIrange{32137.5}{32147.5}{\electronvolt}). The scans are performed in alternating directions (from low to high $qU$ and back) to compensate for a possible slow drift of the source activity over time. The data for each mode are combined by adding the counts and live times for each $qU$ scan step. The model of the N${_{2,3}}$-32 spectrum is fitted to the resulting two measured spectra for each pixel. Pixel-wise analysis is possible in this case, because of high statistics and smaller parameter correlations in the N${_{2,3}}$-32 lines model. The same selection of 126 pixels as for the K-32 line measurement is used.

First, the spectrum taken in the symmetric configuration is analysed. The inferred best-fit parameters are used as reference values in the analysis of the SAP scans of the N lines: the effective position of the N$_2$-32 line, the separation and the relative probability of N$_2$-32 and N$_3$-32 lines, and the angular detection efficiency. The likelihood profiling is performed with a Markov Chain Monte Carlo for these parameters for a robust estimation of uncertainties and correlations, which are used in the analysis of the SAP scans. 
The remaining fit parameters are: the background rate $R_\mathrm{bg}$, $B_\mathrm{ana}$, $\sigma^2$, the intensity of N$_2$-32 line and the column density of tritium gas. The latter is constrained with a pull term with \SI{5}{\percent} systematic uncertainty. The magnetic field $B_\mathrm{ana}$, as opposed to the case of K-32 line, is an unconstrained parameter of the fit due to the smaller uncertainty.
For both lines the natural line width is fixed to 0 and the $\sigma^2$ parameter is limited to the non-negative values only, because no robust extension of the model to the negative range is possible in the case of $\Gamma = 0$ (in the case of $\Gamma > 0$, the negative $\sigma^2$ parameter can be effectively subtracted from the non-zero $\Gamma$).

Following the method of reference measurements, in the second step the spectrum of the N${_{2,3}}$-32 lines measured in SAP mode is analysed using the inputs from the symmetric-mode measurement. The spectra recorded by each detector pixel $i$ are analysed individually using the corresponding separation, relative intensity, and angular detection efficiency from the reference measurement for pixel $i$, constrained by a multivariate pull term.
For the parameters $B_\mathrm{ana}$, $\sigma^2$ and line position the best fit is determined via a gradient-descent minimization of the negative logarithm of the likelihood (analogous to Eq.~\ref{eq:likelihood}), while the uncertainties and correlations are obtained via a Markov Chain Monte Carlo scan of $-2\ln{\mathcal{L}}$ near its minimum. The goodness-of-fit parameter, defined via the minimal $\chi^2$ divided by the number of degrees of freedom (NDoF) as $\chi^2/\mathrm{NDoF}=-2\ln{\mathcal{L}}/\mathrm{NDoF}$, is typically close to 1 for these fits.

\begin{figure*}[h]
  \frame{\includegraphics[width=0.95\textwidth,center]{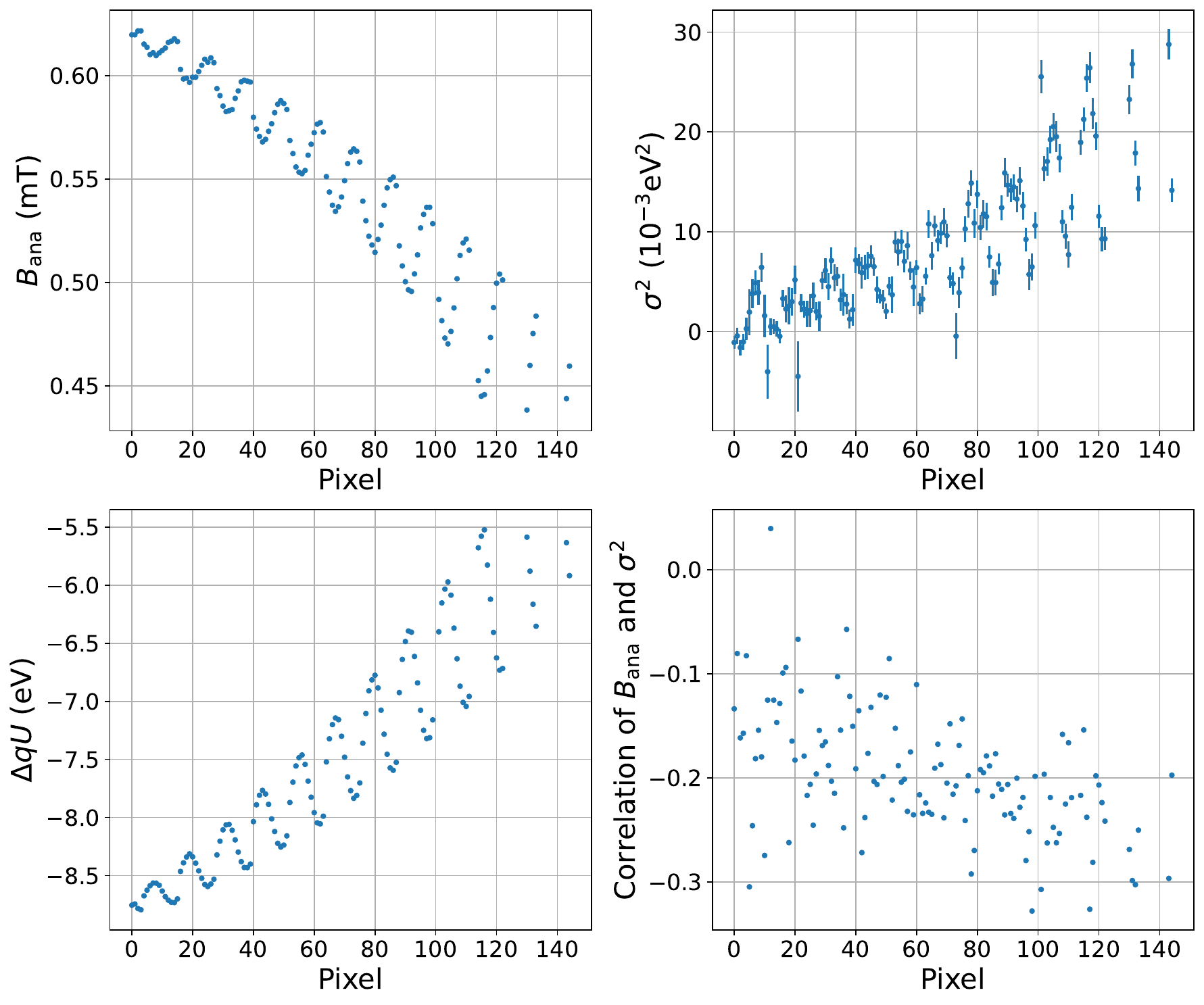}}
\caption{
The magnetic field $B_\mathrm{ana}$, transmission broadening $\sigma^2$, potential depression $\Delta qU$ and the correlation of $B_\mathrm{ana}$ and $\sigma^2$ parameters for the shifted analysing plane. The values are obtained for each detector pixel from the N$_\mathrm{2,3}$-32 lines measurement in SAP configuration, including the constraints from the reference measurement and corrections described in the text. The uncertainties of $B_\mathrm{ana}$ and $\Delta qU$ are too small to be visible.}
\label{fig:9}       
\end{figure*}

Finally, we derive the parameters of the transmission function in the SAP configuration for each detector pixel. The potential in the analysing plane at a given retarding potential $qU$ applied to the spectrometer is calculated from Eqs.~\ref{eq:depression1},\ref{eq:depression2} using the positions of the N$_2$-32 line measured in the two spectrometer configurations and the simulated potential for the symmetrical configuration.
The measured broadening parameter has to be corrected for the broadening within a pixel $\sigma^2_\mathrm{sym,sim}$ for the symmetric configuration and the broadening $\sigma^2_\mathrm{sym,fit}$ measured in the symmetric configuration:
\begin{align}
    \sigma^2_\mathrm{SAP} = \sigma^2_\mathrm{SAP,fit} - \left[ \sigma^2_\mathrm{sym,fit} - \sigma^2_\mathrm{sym,sim} \right].
\label{eq:sigma_correction}
\end{align}
The obtained pixel-wise parameters of the SAP configuration are shown in figure~\ref{fig:9}.

It is important to notice here, that the pixel-wise analysis may yield slightly different best-fit values of the physical parameters (separation of the lines, relative probability, absolute line position). The simultaneous analysis of all 126 pixels with several common and several individual parameters is computationally expensive. We performed a simultaneous fit with common physical parameters to the data of 4 pixels and found a negligible change in the estimates of the parameters of interest ($B_\mathrm{ana}$, $\sigma^2$ and line position) compared to the pixel-wise analysis.

\begin{figure*}[h]
  \frame{\includegraphics[width=0.95\textwidth,center]{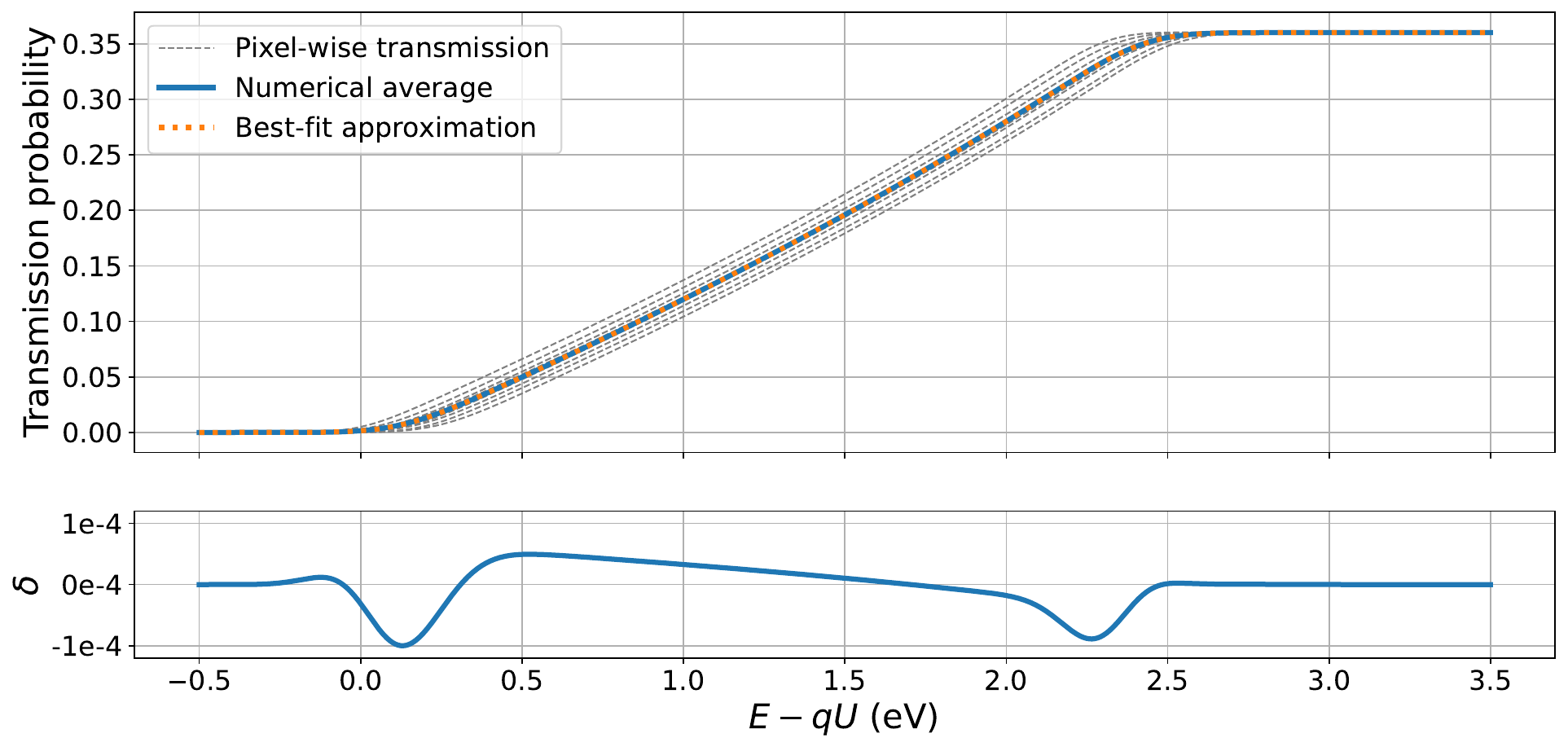}}
\caption{Illustration of the derivation of the patch-wise transmission function parameters. 
Top: Based on the transmission functions for each pixel in the patch (gray, dashed) a numerical average (blue, solid) is calculated. This average is approximated with a single transmission function (orange, dotted). The parameters of the patch transmission function are defined by fitting, the uncertainties are estimated via a Monte Carlo generation.
Bottom: The absolute difference $\delta$ of the numerical average and the approximate transmission functions is shown with the solid blue line. The typical difference $\mathcal{O}(10^{-4})$ can be neglected in the further analysis.}
\label{fig:10}       
\end{figure*}

\textbf{\textit{Deriving patch-wise values for the transmission function in SAP mode.}} 
As mentioned before, the full multi-pixel analysis for the $^\mathrm{83m}$Kr spectra and the neutrino-mass measurements poses significant computational challenges. To facilitate the analysis of the neutrino-mass measurements, it is preferable to reduce the number of calculated model predictions by combining several pixels into patches, as was done for the K-32 line spectrum analysis, and to have one model prediction for each patch. The same pixel combination can be used for the N$_{2,3}$-32 lines analysis. The following averaging procedure was introduced to convert the pixel-wise parameters of transmission into the patch-wise values.
First, the transmission function is calculated for each pixel according to Eq.~\ref{eq:transmission_isotropic} for $qU=\SI{18.6}{\kilo\electronvolt}$ in the range of energies \SIrange[parse-numbers=false]{-0.5}{+3.5}{\electronvolt} around $qU$ with a step of \SI{5E-4}{\electronvolt}. Then the numerical mean of the 9 pixel-wise transmission functions within the respective patch is calculated. Then a single transmission function with an effective magnetic field $B_\mathrm{ana}$, potential $qU_\mathrm{ana}$, and an effective broadening $\sigma^2$ is fitted to the numerical mean. Finally, the uncertainties of the parameters are estimated via a Monte Carlo approach. Several thousand samples are generated for the pixel-wise parameters according to their uncertainties and correlations, then the corresponding numerical average transmissions are fitted as described above. The distribution of the effective parameters of the transmission functions defines the uncertainties of these parameters. The procedure is illustrated in figure~\ref{fig:10}.
It was shown that using these patch-wise or original pixel-wise parameters does not change $m_\nu^2$ estimation on the order of $\mathcal{O}(\SI{0.001}{\electronvolt^2})$, which proves the reliability of the averaging procedure.

The systematic effects are related to the magnetic fields $B_\mathrm{s}$ and $B_\mathrm{max}$. The correlations between the magnetic field and the fit parameters were estimated by adding $B_\mathrm{s}$ and $B_\mathrm{max}$ as constrained fit parameters. The correlation of the transmission parameters to the magnetic field in the source $B_\mathrm{s}$ was found to be negligible.
The correlation to $B_\mathrm{max}$ is somewhat larger in the pixel-wise analysis ($\rho\approx 0.6$), but gets weaker for the patch-wise values ($\rho\approx 0.3$).

\begin{figure*}[h]
  \frame{\includegraphics[width=0.95\textwidth,center]{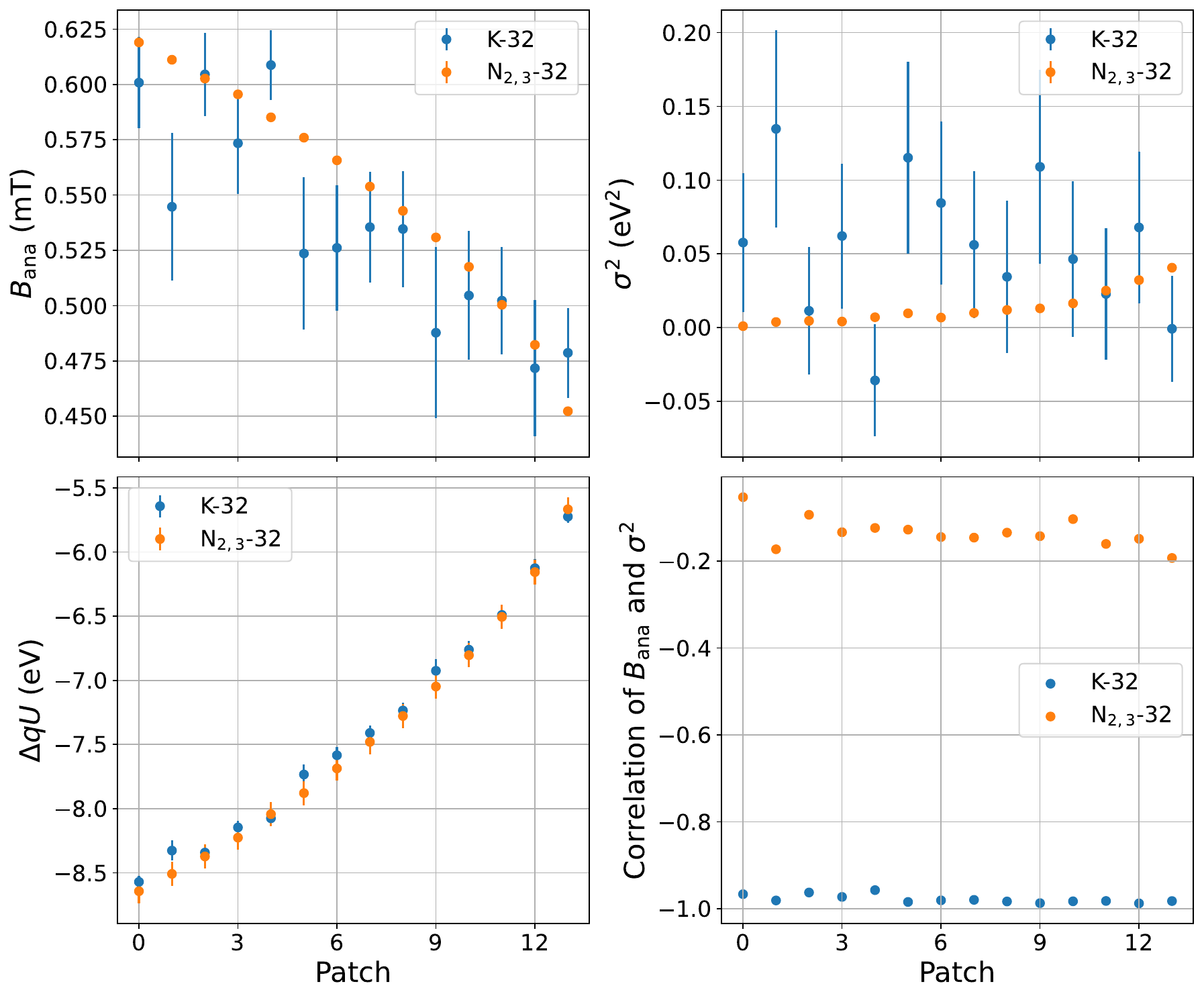}}
\caption{Comparison of the transmission parameters from the K-32 line (blue) and N$_{2,3}$-32 lines (orange) measurements. The magnetic field in the analysing plane is shown in the upper left plot. The magnetic field is expected to get smaller in the outer patches. $B_\mathrm{ana}$ measured with the K-32 line has a high statistical uncertainty and high correlation to the broadening parameter $\sigma^2_\mathrm{ana}$, shown in the upper right plot. The lower left plot shows the potential depression in the SAP configuration; patch-wise values follow the same pattern for both measurements and vary by about \SI{3}{\electronvolt}. The correlation of $B_\mathrm{ana}$ and $\sigma^2_\mathrm{ana}$ is shown in the lower right plot.}
\label{fig:11}       
\end{figure*}

\textbf{\textit{Discussion of the results.}}
The parameters of the SAP configuration were measured using two different calibration lines of $^\mathrm{83m}$Kr and two different source-operation modes. Quantitative comparison is not possible due to a slight modification of the spectrometer configuration between the two measurements (the current of air coil magnet \#11 was changed from \SI{-54}{\ampere} to \SI{-60}{\ampere} due to a modification of the power supply). A summary of the K and N analysis results is shown in figure~\ref{fig:11}. The results of the K-32 line fits show higher statistical fluctuations and significant correlations between the $B_\mathrm{ana}$ and $\sigma^2$ parameters, while the fits of the N$_{2,3}$-32 doublet are more stable. The potential depression $\Delta qU$ varies from \SIrange{-8.5}{-5.5}{\electronvolt}, while the magnetic field $B_\mathrm{ana}$ falls off towards the outer part of the spectrometer from \SIrange{625}{450}{\micro\tesla}. Due to higher inhomogeneity of the fields in the outer parts of the spectrometer, which are mapped onto the outer patches, the broadening increases with the patch number. 

An additional effect is more pronounced in the N$_{2,3}$-32 measurement. It is the change of the potential depression for different absolute energies due to the presence of the ground electrodes at both ends of the spectrometer. The pixel-wise distribution of $qU_\mathrm{ana}$ values is therefore different at \SI{32}{\kilo\electronvolt} and \SI{18.6}{\kilo\electronvolt}.

To take this effect into account, simulations of the SAP configuration were performed to derive the relative change of the transmission parameters from $qU=\SI{32}{\kilo\electronvolt}$ (N$_{2,3}$-32 lines) to $qU=\SI{18.6}{\kilo\electronvolt}$ (tritium endpoint). 
Even though the absolute values of $qU_\mathrm{ana}$, $B_\mathrm{ana}$ and $\sigma^2$ are not fully reliable due to imperfect knowledge of the setup's geometry, their relative change can be used to estimate the effect. The main impact is on the potential, which is shifted by up to \SI{90}{\milli\volt} with a radial variation by \SI{30}{\milli\volt}. The magnetic field $B_\mathrm{ana}$ varies by about \SI{1}{\micro\tesla} and the change of the $\sigma^2$ parameter is below \SI{0.1E-3}{\electronvolt^2}. A conservative estimate of the uncertainty -- the maximum of the correction among the patches -- was chosen, given that the simulations may not be precise enough also for relative change estimation. These uncertainties were added in an uncorrelated way to the uncertainties of the transmission parameters from the N$_{2,3}$-32 measurement.

Additional measurements of the electromagnetic fields in symmetric and shifted analysing-plane configurations were performed using a monoenergetic photo-electron source~\cite{Behrens:2017cmd}, installed at the upstream end of KATRIN beamline. A qualitative agreement of both magnetic-field and electric-potential estimation with the $^\mathrm{83m}$Kr-based measurement was demonstrated, though the measurement-related systematics do not allow for a detailed comparison~\cite{Block2022_1000145073}.

\section{Impact on the neutrino mass analysis}
\label{sec:7}
The reduction of the spectrometer background rate by a factor of 2 with the SAP configuration is improving the statistical sensitivity of the KATRIN experiment to the neutrino mass. However, the overall sensitivity improves only if the SAP-related systematic effects are small enough.

To ensure that the calibration of the electromagnetic fields is sufficiently precise, we estimate the impact of the derived transmission-parameter uncertainties on the neutrino-mass sensitivity of KATRIN.
For that, two sensitivity studies are performed based on the settings of the third and fifth neutrino-mass-measurement campaigns. These two campaigns are chosen because the field calibration measurement with the K-32 line was performed before the third campaign (May/June 2020), and the measurement with the N$_{2,3}$-32 lines was done after the fifth campaign was completed (August 2021), see \cite{Aker:2024drp} for the details of the measurement configurations.

The typical operational parameters of the third neutrino-mass campaign are: 
\begin{itemize}
    \item source density: \SI{40}{\percent} of the nominal one
    \item source temperature: \SI{80}{\kelvin}
    \item spectrometer field configuration: SAP with the parameters measured with K-32 line, before the change of air coil \#11 current
    \item background rate: 0.116~cps (counts per second) for 126 active pixels of the detector
\end{itemize}

The systematic uncertainties due to the electromagnetic fields are introduced with a multivariate pull term on the $B_\mathrm{ana}$ and $\sigma_\mathrm{ana}^2$ parameters for each patch.
The systematic uncertainty on $m_\nu^2$ is estimated as the difference of the total uncertainty to the statistical uncertainty:
$\sigma_{m^2,\mathrm{syst}}^2=\sigma_{m^2,\mathrm{stat+syst}}^2-\sigma_{m^2,\mathrm{stat}}^2$. The resulting value $\sigma_{m^2,\mathrm{syst}}=\SI{0.007}{\electronvolt^2}$ is close to the design requirement of KATRIN. It is almost independent from the overall statistical uncertainty $\sigma_{m^2,\mathrm{stat}}\gg \sigma_{m^2,\mathrm{syst}}$. 

For the estimation of the SAP systematic-uncertainty contribution from the N$_{2,3}$-32 lines measurement, the configuration of the sensitivity study is modified to match the conditions of the fifth neutrino-mass-measurement campaign. The density of the source is increased to \SI{75}{\percent} of the design value, the background is slightly increased to 0.137~cps. The systematic contribution is estimated in the same way as described above and yields $\sigma_{m^2,\mathrm{syst}}<\SI{0.002}{\electronvolt^2}$ which is significantly better than \SI{0.0075}{\electronvolt^2} requirement. The improvement comes from the decorrelation of the $B_\mathrm{ana}$ and $\sigma^2_\mathrm{ana}$ parameters by the sharp N$_{2,3}$-32 lines and the high statistics, available in the krypton and tritium co-circulation mode.
These estimates show that the systematic uncertainty of the shifted analysing plane becomes a small contribution compared to the total uncertainty of the neutrino-mass measurement by KATRIN.

The SAP configuration reduces the statistical uncertainty on $m_\nu^2$ by about \SI{25}{\percent} compared to the nominal symmetric setting, due to a smaller background rate (0.116~cps instead of 0.220~cps in nominal configuration) and similar or better energy resolution. This results in about \SI{12}{\percent} better statistical sensitivity to the neutrino mass $m_\nu$.
Additionally, the SAP configuration reduces the non-Poisson overdispersion of the background counts \cite{Lokhov:2022iag}, and effectively mitigates its impact on the neutrino-mass estimation.

The patch-wise analysis of the neutrino-mass measurements performed in the SAP configuration, compared to the uniform analysis used in the first two campaigns of KATRIN~\cite{KATRIN:2021uub}, requires about an order of magnitude more computations of the $\upbeta$-decay spectrum model. A significant boost of the spectrum calculation speed was achieved via fast predictions using neural networks~\cite{Karl:2022jda} and via extensive code optimizations with caching.

It is worth mentioning that the calibration measurement with $^\mathrm{83m}$Kr has to be repeated after any modification of the setup geometry (e.g. maintenance of the focal plane detector). Then the measured fields always correspond to the ones that are used for the neutrino-mass measurements. However, this calibration procedure does not require significant measurement time ($\mathcal{O}$(1 day)), provided that the $^\mathrm{83m}$Kr-source activity is high enough, and it is performed regularly, 1-2 times per year. The monitoring of the magnetic fields during the neutrino-mass measurements is conducted by precise magnetometers, installed around the main spectrometer~\cite{Letnev:2018fkq}.

\section{Summary and outlook}
\label{sec:8}
The reduction of the background rate in the KATRIN neutrino-mass measurement is one of the key challenges in improving the sensitivity of the experiment. The shifted-analysing-plane configuration was introduced to decrease the background rate by reducing the spectrometer volume between the analysing plane and the detector, while at the same time preserving the spectrometer energy resolution below \SI{2.8}{\electronvolt}. The complex layout of the electromagnetic fields in this new configuration makes the simulations less reliable and requires a direct measurement of the spectrometer transmission properties.

In this work we described the in situ calibration of the electromagnetic fields in the main spectrometer of KATRIN using the conversion electrons of gaseous $^\mathrm{83m}$Kr. The measurement procedure allows a quick calibration, while the main (source-related) systematic effects are cancelled out due to the use of a reference measurement scheme with a well-known spectrometer configuration.

The method was first applied using K-32 electrons. The rather high natural K-32 line width leads to a significant anti-correlation of the key parameters of the transmission function ($B_\mathrm{ana}$ and $\sigma^2_\mathrm{ana}$). To stabilize the fit and facilitate the analysis, the detector pixels are grouped into patches with higher statistics and a single spectrum-model prediction per patch.

After demonstrating the principle of the field measurement, a better precision was achieved by the use of the narrow N$_{2,3}$-32 lines. Here the $B_\mathrm{ana}$ and $\sigma^2_\mathrm{ana}$ parameters become effectively uncorrelated, because they have different impacts on the measured spectral shape. Due to the higher energies of the N$_{2,3}$-32 electrons, the estimation of the retarding potential in the SAP $qU_\mathrm{ana}$ is corrected, when used for the neutrino mass analysis.

The sets of transmission parameters $\left\{ B_\mathrm{ana,i},\sigma^2_\mathrm{ana,i},qU_\mathrm{ana,i} \right\} $ for patches $i=0..13$ are used in the analysis of the neutrino-mass measurements performed in the shifted-analysing-plane configuration, including the recent release of KATRIN neutrino-mass bound of \SI{0.45}{\electronvolt} at 90\,\% CL~\cite{Aker:2024drp}.
The systematic contribution from the SAP fields to the neutrino mass uncertainty is $\SI{0.007}{\electronvolt^2}$ for the K-32 line measurement and below $\SI{0.002}{\electronvolt^2}$ for the N$_{2,3}$-32 doublet data. Both satisfy the requirements to the systematic uncertainties of KATRIN.
The calibration measurements are performed regularly to ensure a good control of the fields in the main spectrometer over the course of neutrino-mass measurements in the SAP mode.
The precise determination of the magnetic field in the symmetric configurations with the N$_{2,3}$-32 lines are also used for inputs of the source column density determination with the photo-electron source.

The overall improvement of the neutrino-mass sensitivity of KATRIN due to the background-rate reduction in SAP configuration significantly outweighs the small additional systematic uncertainties of the transmission parameters, measured via the method presented in this paper. Therefore the SAP configuration became the default mode for neutrino-mass measurements at KATRIN since 2020.

\begin{acknowledgements}

We acknowledge the support of Helmholtz Association (HGF), Ministry for Education and Research BMBF (05A23PMA, 05A23PX2, 05A23VK2, and 05A23WO6), the doctoral school KSETA at KIT, Helmholtz Initiative and Networking Fund (grant agreement W2/W3-118), Max Planck Research Group (MaxPlanck@TUM), and Deutsche Forschungsgemeinschaft DFG (GRK 2149 and SFB-1258 and under Germany's Excellence Strategy EXC 2094 – 390783311) in Germany; Ministry of Education, Youth and Sport (CANAM-LM2015056) in the Czech Republic; Istituto Nazionale di Fisica Nucleare (INFN) in Italy; the National Science, Research and Innovation Fund via the Program Management Unit for Human Resources \& Institutional Development, Research and Innovation (grant B37G660014) in Thailand; and the Department of Energy through grants DE-FG02-97ER41020, DE-FG02-94ER40818, DE-SC0004036, DE-FG02-97ER41033, DE-FG02-97ER41041,  {DE-SC0011091 and DE-SC0019304 and the Federal Prime Agreement DE-AC02-05CH11231} in the United States. This project has received funding from the European Research Council (ERC) under the European Union Horizon 2020 research and innovation programme (grant agreement No. 852845). We thank the computing cluster support at the Institute for Astroparticle Physics at Karlsruhe Institute of Technology, Max Planck Computing and Data Facility (MPCDF), and the National Energy Research Scientific Computing Center (NERSC) at Lawrence Berkeley National Laboratory.

\end{acknowledgements}

\bibliographystyle{spphys}       
\bibliography{bibliography}   

\end{document}